\documentclass[letterpaper, 10 pt, conference]{ieeeconf}
\IEEEoverridecommandlockouts
\usepackage{cite}
\usepackage{amsmath,amssymb,amsfonts}
\usepackage{algorithmic}
\usepackage{graphicx}
\usepackage{textcomp}
\usepackage{xcolor}
\usepackage{bm}
\title{\LARGE \bf
Efficient Learning of Vehicle Controller Parameters via Multi-Fidelity Bayesian Optimization: From Simulation to Experiment
}

\author{Yongpeng Zhao$^{1,2}$, Maik Pfefferkorn$^{2}$, Maximilian Templer$^{1}$ and Rolf Findeisen$^{2}$
\thanks{This work was supported by Volkswagen AG.}
\thanks{$^{1}$Yongpeng Zhao and Maximilian Templer are with Group Innovation, Volkswagen AG, Wolfsburg, Germany {\tt\small yongpeng.zhao@volkswagen.de, maximilian.templer1@volkswagen.de}}%
\thanks{$^{2}$Yongpeng Zhao, Maik Pfefferkorn and Rolf Findeisen are with the Control and Cyber-Physical Systems Laboratory, Technical University of Darmstadt, Darmstadt, Germany {\tt\small maik.pfefferkorn@iat.tu-darmstadt.de, rolf.findeisen@iat.tu-darmstadt.de}}%
}
\begin{document}
	\maketitle
	\thispagestyle{empty}
    \pagestyle{empty}
	\begin{abstract}
        Parameter tuning for vehicle controllers remains a costly and time-intensive challenge in automotive development. Traditional approaches rely on extensive real-world testing, making the process inefficient. We propose a multi-fidelity Bayesian optimization approach that efficiently learns optimal controller parameters by leveraging both low-fidelity simulation data and a very limited number of real-world experiments. Our approach significantly reduces the need for manual tuning and expensive field testing while maintaining the standard two-stage development workflow used in industry. The core contribution is the integration of an auto-regressive multi-fidelity Gaussian process model into Bayesian optimization, enabling knowledge transfer between different fidelity levels without requiring additional low-fidelity evaluations during real-world testing. We validate our approach through both simulation studies and real-world experiments. The results demonstrate that our method achieves high-quality controller performance with only very few real-world experiments, highlighting its potential as a practical and scalable solution for intelligent vehicle control tuning in industrial applications.
	\end{abstract}
	
	\section{Introduction}
	\label{Intro}
	In recent years, the automobile industry has been rapidly evolving, requiring shorter development cycles and higher performance standards for vehicle controllers. Products with enhanced functionalities are being developed within increasingly shorter time frames~\cite{llopis2021impact}, posing considerable challenges to current development processes, especially the development of vehicle controllers~\cite{gusikhin2008intelligent}. The performance of these controllers depends on their structure but also on a set of carefully tuned parameters, which play a critical role in ensuring optimal yet safe behavior under real-world driving conditions. Consequently, the V-model development process, widely adopted in the automotive industry, divides controller development into two stages: architecture definition and system integration~\cite{GraesslerHentze}. 
    In the architecture definition stage, the overall controller structure is designed, and simplified simulation models are typically used to evaluate feasibility and performance. 
    However, these models often fail to capture the full complexity of real vehicle dynamics, introducing inaccuracies that necessitate additional fine-tuning~\cite{smith1995}. 
    During the system integration stage, the controller is tested on a real vehicle, where application engineers manually tune the parameters to achieve the desired closed-loop performance. 
    This manual tuning process is both time-consuming and costly, requiring extensive real-world testing and expert knowledge. 
    Furthermore, achieving consistent brand-specific system behavior as well as satisfactory objective and subjective performance evaluation is a significant challenge~\cite{Auckland2008}.
    A key limitation of the current approach is its heavy reliance on real-world experiments, which are expensive and difficult to scale. While simulation-based tuning is faster and more convenient, it does not fully reflect real-world conditions, often leading to suboptimal parameter choices. This raises a fundamental question: how can we systematically reduce the dependence on real-world experiments while ensuring satisfactory and robust performance?
	
	The challenge of controller tuning is a well-recognized issue. Classical methods, such as the Ziegler-Nichols approach~\cite{ziegler1942optimum}, primarily target PID controllers with simple structures, offering limited flexibility for complex controllers. 
    Iterative feedback tuning~\cite{hjalmarsson1998iterative} and extremum seeking~\cite{Killingsworth2006} impose weaker assumptions on the controller architecture but may fail due to their reliance on local optimization. Meta-heuristic methods, such as genetic algorithms~\cite{Porter1992GeneticTO} and particle swarm optimization~\cite{1300705}, can explore global optima but typically require an excessively large number of data samples, rendering these methods inapplicable in combination with costly experiments. Recently, Bayesian optimization (BO) has gained prominence as an effective and sample-efficient approach for automatic controller learning~\cite{neumann2019data,Holzmann2024,Hirt2024a,hirt2024time,Hirt2025}. BO offers several key advantages: it is a global black-box optimization method, demonstrates high sample-efficiency, and naturally accounts for noise in the surrogate modeling process~\cite{lizotte2007automatic}.
	
	Although standard BO has proven highly effective for learning parameters of control systems, collecting real-world data is often prohibitively expensive and time-consuming. To address this challenge, various solutions have been proposed to further reduce the data requirements for BO. One widely adopted approach is multi-fidelity BO~\cite{kandasamy2016gaussian}: it leverages information from multiple fidelity levels and minimizes the dependence on real-world data. For instance,~\cite{marco2017virtual} extended entropy search to incorporate multiple information sources, thereby improving the cost efficiency of standard BO. However, like other multi-fidelity BO methods~\cite{wang2023recent}, this approach relies on continuous low-fidelity evaluations and experiments throughout the learning process, which contradicts the conventional vehicle controller development workflow.
	
	We employ BO for the efficient learning of optimal parameters considering the case of tuning a trajectory tracking controller. The key contribution of this work lies in leveraging existing low-fidelity data from earlier design and tuning phases such as from simulations to accelerate the optimization process in real-world environments. This is achieved through multi-fidelity surrogate modeling using a linear auto-regressive Gaussian process. Notably, our proposed method does not require additional evaluations of the objective function at low-fidelity levels during the optimization, which preserves the conventional development process of vehicle controllers. 
    We demonstrate the effectiveness of the proposed approach considering the combined simulation-based and experimental controller parameter tuning of a test vehicle.
	
	The remainder of this article is organized as follows. Section~\ref{Sec II} outlines the problem formulation, introducing the considered system and controller structure. 
    Section~\ref{Sec III} provides an overview of Bayesian optimization, including single- and multi-fidelity Gaussian process regression, and details the problem of controller parameter learning. 
    In Section~\ref{Simulation Studies}, we tune the controller parameters in simulations, comparing our approach with alternative methods. Section~\ref{Experimental Studies} considers the refinement of the control parameters on a test vehicle in real-world environments, before concluding in Section~\ref{CaO}.
	
	\section{Problem Formulation}
	\label{Sec II}
\begin{figure}[tb]
		\centerline{\includegraphics{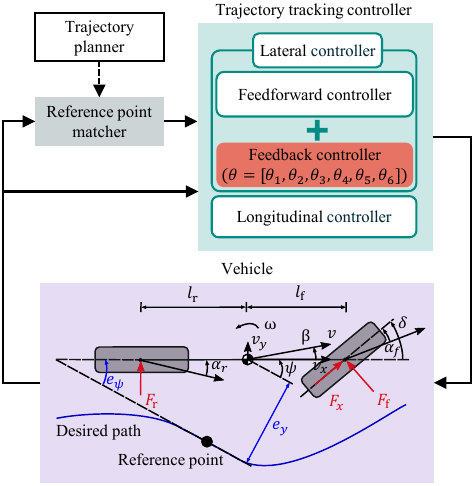}}
		\caption{Architecture of the trajectory tracking control system.}
		\label{Fig-architecture}
	\end{figure}

     First, we provide a high-level overview of the problem formulation. We present our approach for controller parameter tuning in the context of tuning a trajectory tracking controller, as illustrated in Fig.~\ref{Fig-architecture}. A vehicle model that captures the key dynamics relevant to trajectory tracking is available and is utilized within the two-stage development process of the considered controller. After designing the control architecture, the controller parameters must be tuned and adjusted to the specific vehicle and operating conditions. We propose a two-stage procedure to achieve this: simulation-based parameter tuning, followed by parameter refinement through real-world experiments. Instead of manual tuning, we employ BO in both stages and employ an auto-regressive multi-fidelity Gaussian process as the surrogate model in the second stage. This approach exploits knowledge from simulations to improve efficiency in real-world experiments, thereby reducing the need for extensive field testing. In the following, we describe the considered vehicle model and control structure, before outlining the multi-fidelity parameter learning strategy in Section~\ref{Sec III}.

    The relevant vehicle dynamics for trajectory tracking control are described by the continuous-time model
	\begin{equation}
		\label{eq-vehicle model} 
		\dot{x}=f(x,u),
	\end{equation}
	with $x \in \mathbb{R}^{n_\mathrm{x}}$ as the state, $u \in \mathbb{R}^{n_\mathrm{u}}$ as the control input, and the dynamics $f:\mathbb{R}^{n_\mathrm{x}} \times \mathbb{R}^{n_\mathrm{u}} \to \mathbb{R}^{n_\mathrm{x}}$. The vehicle state is composed of $x=[p_x,p_y,\psi,v_x,v_y,\omega]$, where $p_x,p_y,\psi$ denote the position and heading angle of the vehicle in the global coordinate frame and $v_x,v_y,\omega$ are the velocities and yaw rate defined in the vehicle's body frame. The control input is comprised of $u=[\delta,\tau]$, with $\delta$ as the steering angle and $\tau$ as the drive train command. The dynamics $f$ is governed by a simple single track model
	\begin{equation}
		\label{eq-bicycleModel}
		\begin{bmatrix}
            \dot{p}_x \\
            \dot{p}_y \\
            \dot{\psi} \\
			\dot{v}_x \\
			\dot{v}_y \\
			\dot{\omega}
		\end{bmatrix}
		=
		\begin{bmatrix}
            v_x \\
            v_y \\
            \omega \\
			\frac{1}{m}(F_x\cos(\delta)-F_\mathrm{f}\sin(\delta)+mv_y\omega) \\
			\frac{1}{m}(F_x\sin(\delta)+F_\mathrm{f}\cos(\delta)+F_\mathrm{r}-mv_x\omega) \\
			\frac{1}{I_z}((F_x\sin(\delta)+F_\mathrm{f}\cos(\delta))l_\mathrm{f}-F_\mathrm{r}l_\mathrm{r})
		\end{bmatrix},
	\end{equation}
	with $m$ as the vehicle's mass, $I_z$ as the moment of inertia, and $l_{\mathrm{f}}$ and $l_{\mathrm{r}}$ as the distance between the center of gravity and the front and rear axle respectively. $F_x$ denotes the longitudinal tire force, which depends on $\tau$. The lateral tire forces $F_{\mathrm{f}}$ and $F_{\mathrm{r}}$ are calculated via the simplified magic formula~\cite{pacejka2005tire}
	\begin{equation}
		\label{eq-mfTire}
		F_{\mathrm{f,r}}=D_{\mathrm{f,r}}\sin(C_{\mathrm{f,r}}\arctan(B_{\mathrm{f,r}}\alpha_{\mathrm{f,r}})),
	\end{equation}
	where the parameters $B_{\mathrm{f,r}}$, $C_{\mathrm{f,r}}$ and $D_{\mathrm{f,r}}$ are commonly identified from data. The slip angles $\alpha_{\mathrm{f,r}}$ are calculated via
	\begin{equation}
		\label{eq-slipAngle}
		\alpha_{\mathrm{f}} \! = \! \delta  -  \arctan \left( \! \frac{v_y \! + \! \omega l_{\mathrm{f}}}{v_x} \! \right), ~
		\alpha_{\mathrm{r}} \! = \! -\arctan \left( \! \frac{v_y \! - \! \omega l_{\mathrm{r}}}{v_x} \! \right) \!.
	\end{equation}
    We note that the parameters introduce uncertainty and will differ from reality depending on the environmental conditions, as well as the vehicle design.
    
	The objective is to control system \eqref{eq-vehicle model} to optimally follow a reference trajectory. To this end, we employ the control structure shown in Fig.~\ref{Fig-architecture}, which is composed of a trajectory planner, a state observer, and a tracking controller. 
    In this work, trajectory planning and tracking control are structurally separated. Specifically, the trajectory planner operates offline and provides reference trajectories in the state space for a given test track. Based on the reference trajectory and current vehicle states, a reference point matcher is repeatedly executed during tracking to generate a reference point and the corresponding reference states by interpolating the waypoints from the planner. In our concrete example, we employ the planner and matcher implementation from~\cite{gundlach2020zeitoptimale}. The reference states are tracked by the tracking controller, which is comprised of one controller for the longitudinal vehicle dynamics and one controller for the lateral vehicle dynamics, producing $\delta$ and $\tau$ respectively. Further details about the control system are provided in~\cite{templerAachen}.  
    The longitudinal controller, which tracks a reference speed profile, is typically uncritical in view of tuning, and its parameters can usually be tuned manually within a reasonable time. In practice, tuning the lateral controller typically poses a greater challenge. Hence, in the scope of this work, we focus on the lateral controller which provides the steering angle $\delta = \delta_{\mathrm{FFW}} + \delta_{\mathrm{FB}}$ and is comprised of a feedforward part, $\delta_\mathrm{FFW}$, and a feedback part, $\delta_\mathrm{FB}$.
	
	The feedback controller is a PID-based controller, which is parameterized by six tuning parameters $\theta = (\theta_1, \theta_2, \theta_3, \theta_4, \theta_5, \theta_6) \in \mathbb{R}^6$ and consists of multiple control loops. 
    These parameters include the coefficients for the proportional, integral, and derivative terms, as well as the time constants for first-order output filters. 
    They need to be carefully tuned in order to achieve a desired closed-loop performance. 
    Additionally, the controller incorporates heuristic rules to manage the complexity of vehicle trajectory tracking, deviating from the standard PID structure. 
    Consequently, classical PID tuning methods, such as the Ziegler-Nichols approach, become impracticable. 
    We aim for learning optimal controller parameters automatically and efficiently using closed-loop data from simulations and real-world experiments. 
    To this end, we employ a multi-fidelity BO scheme which provides enhanced sample efficiency regarding the real-world data compared to standard BO.
	
	\section{Parameter Learning via Bayesian Optimization}
	\label{Sec III}
	In this section, we first introduce the fundamentals of Gaussian process regression, including an overview of auto-regressive multi-fidelity Gaussian processes. Subsequently, we introduce the basics of BO, relying on Gaussian process surrogate models, and conclude this section with some details of learning optimal controller parameters.
	\subsection{Basics of Gaussian Process Regression}\label{subsection-BoGPR}
    Gaussian process (GP) regression provides a probabilistic framework for modeling unknown functions from data, resulting in probabilistic models that provide not only predictions of unknown function values but also uncertainty estimates.
    Their probabilistic nature is key for employing GP surrogate models in Bayesian optimization. 
    
    A Gaussian process \cite{williams2006gaussian}, denoted by
	\begin{equation}
		g(\xi) \sim \mathcal{GP}(\mu(\xi),k(\xi,\xi')),\label{eq-GP}
	\end{equation}
	is defined as a collection of Gaussian random variables $g(\xi)$, indexed by $\xi$, any finite number of which have joint and consistent Gaussian distribution~\cite{williams2006gaussian}. A GP \eqref{eq-GP} is fully defined by its mean function $\mu:\mathbb{R}^{n_\xi} \to \mathbb{R}, \xi \mapsto \mathbb{E}[g(\xi)]$ and its covariance function $k:\mathbb{R}^{n_\xi} \times \mathbb{R}^{n_\xi} \to \mathbb{R}, (\xi, \xi') \mapsto \mathbb{C} \mathrm{ov}[g(\xi), g(\xi')]$.
	
	We use a GP \eqref{eq-GP} to model an unknown function $\varphi:\mathbb{R}^{n_{\xi}} \to \mathbb{R},\xi \mapsto \varphi(\xi)$. Particularly, for each $\xi$ in $\mathbb{R}^{n_{\xi}}$, the Gaussian random variable $g(\xi)$ models the function value $\varphi(\xi)$. The objective is to predict an unobserved function value $\varphi(\xi_*)$ at an arbitrary test location $\xi_*$. To this end, we rely on a set of (potentially noisy) observations of the unknown function, given by $\mathcal{D}=\{ \xi_i, \varphi(\xi_i)+\varepsilon_i \mid i=1, \ldots, m \}$. Therein, $\varepsilon_i \in \mathcal{N}(0,\sigma_n^2)$ models independent and identically distributed white Gaussian noise with variance $\sigma_n^2$. For brevity, we denote the so-called training input matrix by $\Xi \in \mathbb{R}^{m \times n_{\xi}}, ~\Xi_{i:} = \xi_i^\top$, and the training target vector by $\gamma \in \mathbb{R}^m, ~\gamma_i=\varphi(\xi_i)+\varepsilon_i$.
	
	We now consider the joint distribution of the random vector $g(\Xi)=(g(\xi_1),\ldots,g(\xi_m))$ and the random variable $g(\xi_*)$, which is by definition defined via the GP model \eqref{eq-GP}. Noting that $\gamma$ is a realization of $g(\Xi)$, we can compute the posterior (or predictive) distribution of $g(\xi_*)$ by conditioning on the data as $g(\xi_*) \mid \Xi,\gamma \sim \mathcal{N}(\mu^+(\xi_*), k^+(\xi_*, \xi_*))$, with
	\begin{subequations}
		\begin{align}
			\mu^+(\xi_*) &= \mu(\xi_*) + k(\xi_*,\Xi)k_{\gamma}^{-1}(\gamma-\mu(\Xi)) \label{eq-GP-mean} \\
			k^+(\xi_*) &= k(\xi_*,\xi_*) - k(\xi_*,\Xi)k_{\gamma}^{-1}k(\Xi,\xi_*), \label{eq-GP-covar}
		\end{align}
	\end{subequations}
	where $k_{\gamma}=k(\Xi,\Xi)+\sigma^2_n\mathbb{I}$. The posterior mean \eqref{eq-GP-mean} is an estimate for $\varphi(\xi_*)$, while the posterior variance \eqref{eq-GP-covar} quantifies the prediction uncertainty. 
    
    The prior mean function $\mu$ and the prior covariance function $k$ are design choices, that often depend on a set of free hyperparameters. To obtain a meaningful prediction model, these hyperparameters need to be carefully adjusted to the problem. One way to do so is to infer suitable hyperparameters from the training data $\mathcal{D}$ via evidence maximization. Evidence maximization corresponds to adjusting the model parameters such that the likelihood of the training observations under the model is maximized. For details, we refer the reader to~\cite{williams2006gaussian}.

	\subsection{Multi-Fidelity Gaussian Process Regression}\label{subsection-MFGPR}
	Since direct learning from real-world data is expensive, we leverage low-fidelity simulations to reduce the need for costly real-world tests. This requires integrating data from multiple data sources with different fidelity levels systematically, for which dedicated extensions of the standard Gaussian process regression exist. Here, we focus on the so-called AR1 model, first proposed by Kennedy and O'Hagen~\cite{kennedy2000predicting}, which assigns a GP prior to each fidelity level and assumes a linear dependency between the fidelity levels. The detailed formulation of the AR1 scheme is presented in~\cite{kennedy2000predicting,le2014recursive}, while we provide a concise overview in the following. 
	
	
	Consider a system with $s$ levels of information sources denoted as $(\varphi_h(\xi))_{h=1,\ldots,s}$, each of which produces outputs $\gamma_h(\xi)$, and where the level of fidelity increases with $h$. All $s$ levels are modeled by individual GPs $(g_h(\xi))_{h=1,\ldots,s}$ of the form \eqref{eq-GP}. 

	The GP models $(g_h(\xi))_{h=2,\ldots,s}$ of the AR1 scheme are recursively defined by
	\begin{equation}
		\label{eq-AR1-1}
		\left\{
		\begin{array}{l}
			g_h(\xi)={\rho_{h-1}}g_{h-1}(\xi)+\delta_h(\xi), \\
			g_{h-1}(\xi)	\perp \delta_h(\xi), \\
		\end{array}
		\right.
	\end{equation}
	where
	\begin{equation}
		\label{eq-AR1-2}
		\delta_h(\xi) \sim \mathcal{GP}(\mu_h(\xi), k_h(\xi,\xi')),
	\end{equation}
	and
	\begin{equation}
		\label{eq-AR1-3}
		g_1(\xi) \sim \mathcal{GP}(\mu_1(\xi), k_1(\xi,\xi')).
	\end{equation}
	Here, $\rho_{h-1} \in \mathbb{R}$ are scaling factors, $\delta_{h}$ represent additive bias functions, and $\perp$ denotes the independence relationship. It is worth noting that the formulation in \eqref{eq-AR1-1} results in GPs for $(g_h(\xi))_{h=2,\ldots,s}$, since GPs are closed under addition and linear transformation. The hyperparameters of $(g_h(\xi))_{h=1,\ldots,s}$ are inferred via evidence maximization, similarly as in standard Gaussian process regression introduced in Section~\ref{subsection-BoGPR}.    
	A nonlinear version of the AR1 model, known as NARGP, was proposed in~\cite{perdikaris2017nonlinear}, and is defined via
	\begin{equation}
		\left\{
		\begin{array}{l}
			g_h(\xi)=z_{h-1}(g_{h-1}(\xi))+\delta_h(\xi),\label{eq-NARGP-mean}\\
			g_{h-1}(\xi)	\perp \delta_h(\xi), \\
		\end{array}
		\right.
	\end{equation}
	where
	\begin{equation}
		\label{eq-NARGP-2}
		\delta_h(\xi) \sim \mathcal{GP}(\mu_h(\xi), k_h(\xi,\xi')),
	\end{equation}
	and
	\begin{equation}
		\label{eq-NARGP-3}
		g_1(\xi) \sim \mathcal{GP}(\mu_1(\xi), k_1(\xi,\xi')).
	\end{equation} 
    Here, $z_{h-1}: \mathbb{R} \to \mathbb{R}, g(\xi) \mapsto z_{h-1}(g(\xi))$ denote unknown functions, which are analogously modeled using a GP each.
    
	\subsection{Bayesian optimization}
	\label{subsection-BO}
    BO has emerged as a promising approach for solving optimization problems with black-box, expensive-to-evaluate cost functions, 
	\begin{equation}
		\xi^{\star}=\arg \min_{\xi \in \Omega}G(\xi),
	\end{equation}
	where $\xi^{\star}$ is the global minimizer of $G:\mathbb{R}^{n_\xi} \to \mathbb{R},\xi \mapsto G(\xi)$ on the set $\Omega$.
    Unlike more traditional approaches for black-box optimization, which require exhaustive cost function evaluations, BO sequentially selects the most useful locations to evaluate the cost function at, and, thereby, significantly reduces the number of required function evaluations.
    Since the cost function is unknown, BO relies on a surrogate model of $G$, for which GPs are a common choice. During the execution of BO, this surrogate model is sequentially refined in an iterative manner. In particular, in any iteration $n \in \mathbb{N}$, we perform two steps:
	\begin{enumerate}
		\item select a query point $\xi^q_n$ and evaluate the cost function to generate a new data point $(\xi_n^q, G(\xi_n^q))$, and
		\item update the GP model of $G$ using the augmented data set $\mathcal{D}_{n+1}=\mathcal{D}_{n} \cup (\xi_n^q, G(\xi_n^q))$
	\end{enumerate}
	
	To effectively guide the selection of the query points $\xi^q_n$ towards the (global) optimizer $\xi^*$, BO makes use of an acquisition function $\alpha: \mathbb{R}^{n_\xi} \to \mathbb{R}, \xi \mapsto \alpha(\xi; \mathcal{D}_n)$. The acquisition function exploits the GP surrogate model to assess the utility of a query point, trading off exploration of the search space and exploitation of the current best value. In each iteration $n$, the query point is then determined as
	\begin{equation}
		\xi_n^q=\arg \max_{\xi \in \Omega} \alpha(\xi; \mathcal{D}_n).
	\end{equation}
	%
	We employ the so-called expected improvement acquisition function, which aims to maximize the potential improvement over the best observed value of the cost function so far. 
    We refer the reader to \cite{garnett2023bayesian} for a more comprehensive overview of Bayesian optimization.
	
	\subsection{Learning Optimal Controller Parameters}
	\label{subsection OLCP}
	We aim to learn the optimal parameters $\theta^\star$ (in our case those of the trajectory tracking controller), as detailed in Section~\ref{Sec II}, to achieve optimal tracking performance $J(\theta^\star)$ on a test track. 
    To this end, we apply 
    BO, where now $\xi \equiv \theta$ represents the controller parameters and $J$ is the achieved closed-loop performance. In our concrete case, the cost function $J$ is defined as

	\begin{equation}
		\label{eq-J}
		J(\theta)=w_1e_{y,\mathrm{RMS}}+w_2e_{\psi,\mathrm{RMS}}+w_3\ddot{\delta}_{\mathrm{RMS}},
	\end{equation}
	with $e_y$ as the lateral position error, $e_{\psi}$ as the heading angle error, and $\ddot{\delta}$ as the second derivative of the steering angle, c.f., Fig.~\ref{Fig-architecture}. $\mathrm{RMS}$ indicates the root mean square, which is for a time series $q$ given via
	\begin{equation}
		\label{eq-RMS}
		q_{\mathrm{RMS}}=\sqrt{\frac{1}{N_k}\sum_{k=1}^{N_k}(q(t_k))^2},
	\end{equation}
	where $N_k$ represents the number of time steps required to complete a full lap. The weights $[w_1, w_2, w_3]$ are design choices and here set to [1, 3, 0.03], reflecting our preferences in the trade-offs between the cost terms. The cost function \eqref{eq-J} depends on the controller parameters $\theta$ in a complex manner via the quantities $e_y$, $e_{\psi}$, and $\ddot{\delta}$. Note that, there is no closed-form expression available for $J$, which, together with its complexity, motivates the application of BO.

    While standard BO approaches rely on single-fidelity GP surrogate models of $J$ as introduced in Section~\ref{subsection-BoGPR}, we propose modeling the cost function $J$ using the AR1 scheme presented in Section \ref{subsection-MFGPR}.
    In this work, we consider two fidelity levels, $s=2$ and $h \in \{ 1,2 \}$, representing low- and high-fidelity respectively.
    However, although we construct the GP surrogate model from multi-fidelity data, we do not simultaneously acquire data for both the low- and the high-fidelity stage, and hence do not simultaneously update both the low- and high-fidelity component models.
    Instead, we generate low-fidelity data in a first step in simulation, focusing on identifying regions with low, and hence potentially optimal, cost values employing standard BO.
    Thereafter, we add the high-fidelity component to the GP surrogate model and perform BO using real-world experiments with the system in a second step.
    During these real-world experiments, the cost function \eqref{eq-J} is only evaluated at the highest fidelity level, and we do not generate corresponding low-fidelity data by running simulations again.
    Hence, the low-fidelity data set of the AR1 model remains unaltered, while only the high-fidelity data set is updated during the BO iterations.
    Compared to other multi-fidelity BO approaches, where the algorithm has access to the cost function evaluations at all fidelity levels and can alternate between them, our implementation leverages low-fidelity simulation data to efficiently learn the true cost function from fewer real-world samples in a two-step process.
    This aligns with the typical two-stage development process of vehicle controllers.
	
	\section{Simulation Results}\label{Simulation Studies}
\begin{figure}[tb]
		\centerline{\includegraphics{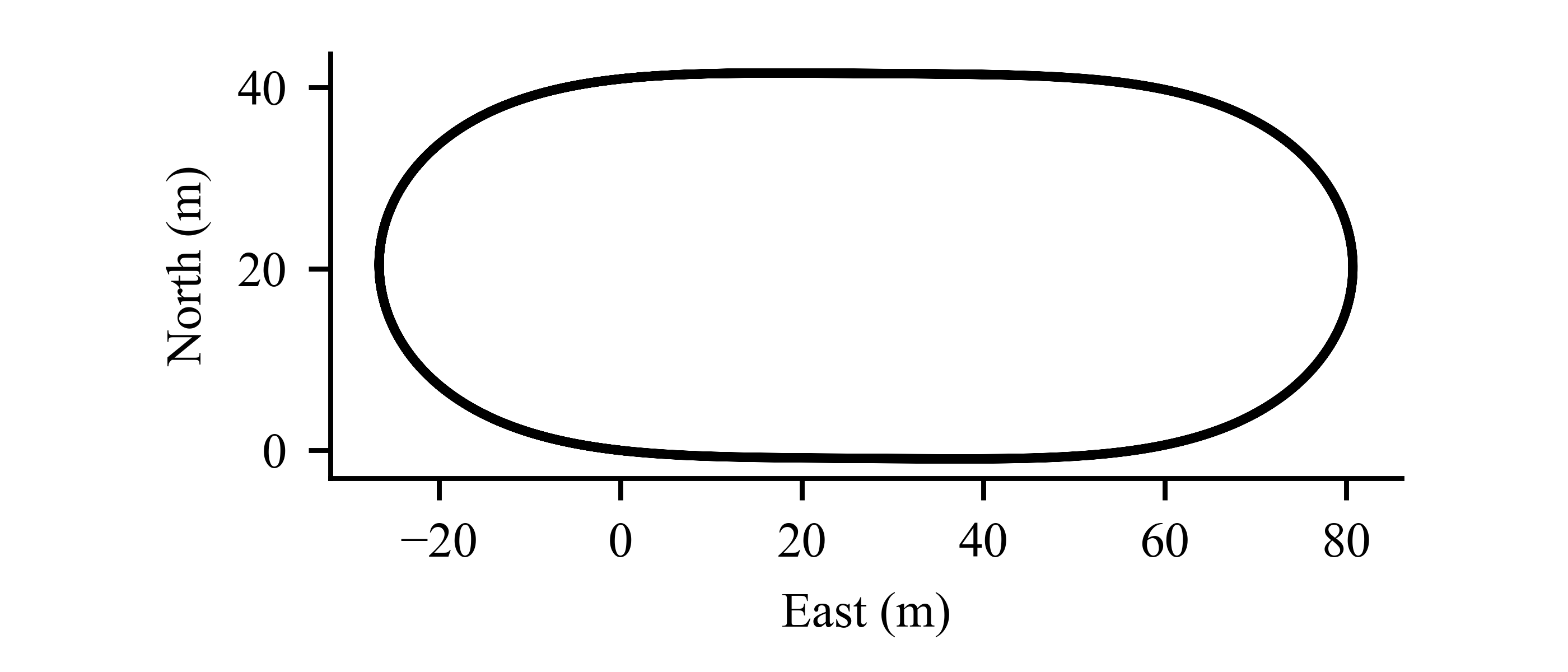}}
		\caption{Reference path on the oval test track.}
		\label{fig-oval}
	\end{figure}

	For simplicity, an oval test track is selected for evaluating the proposed approach in simulations and experiments. The reference trajectory, which is generated by the trajectory planner for the shortest lap time, is shown in Fig.~\ref{fig-oval}.
    \begin{figure}[tb]
		\centerline{\includegraphics{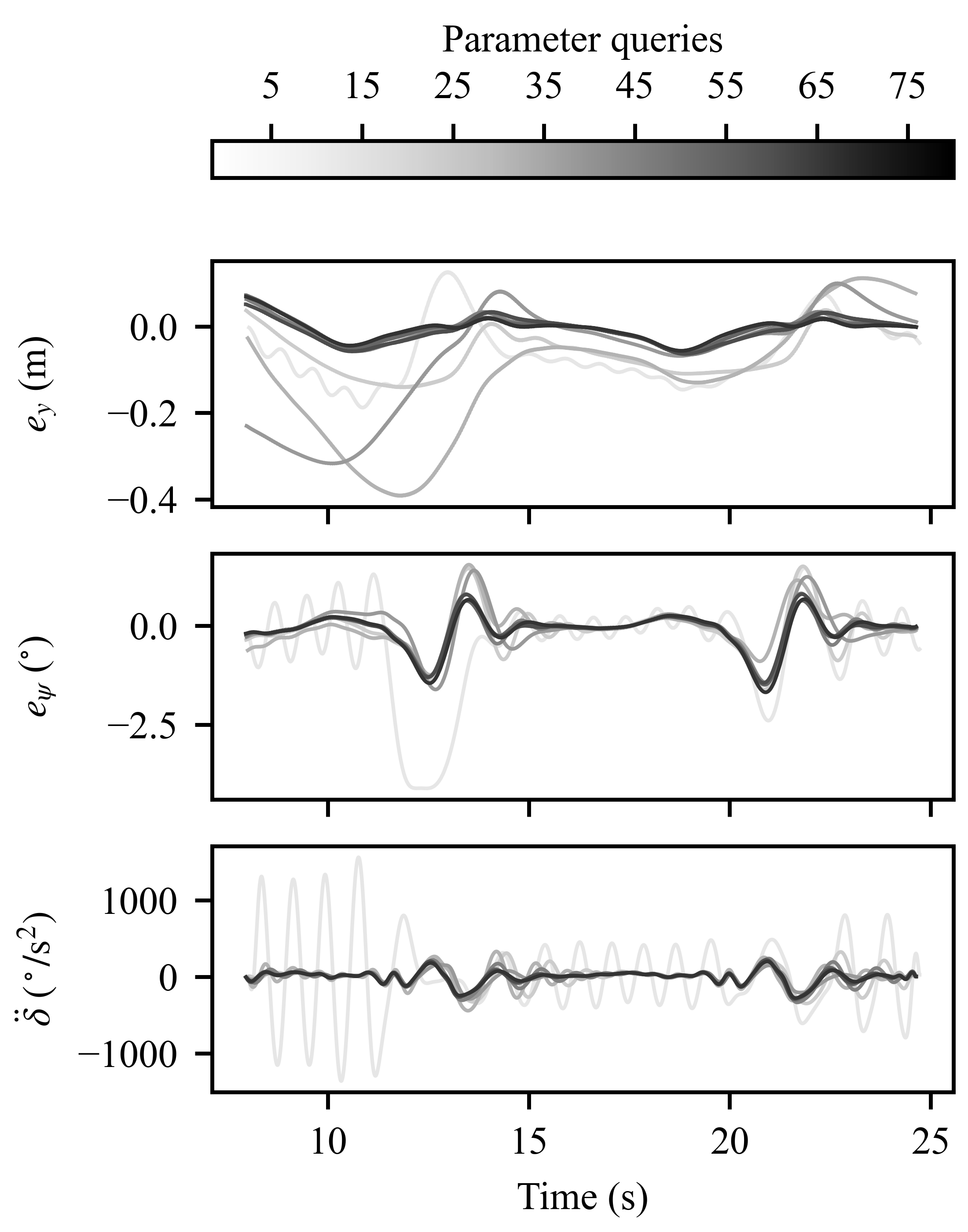}}
		\caption{Trajectory samples of cost terms during the learning process. Darker colors indicate parameter queries evaluated later.}
		\label{fig-tuning stages}
	\end{figure}
	If the autonomous vehicle deviates from the reference path by more than $3\mathrm{m}$, the controller is considered incapable of completing the lap, and the evaluation is terminated. In such cases, a fixed heuristic cost of $J = 0.5$ is assigned, which is conservatively higher than the costs associated with controllers that lead to successful completion of the lap. 
	
    In this section, we showcase the application of the proposed approach in simulations.
    To this end, we discuss the set-up as well as the generation of low-fidelity data in Section \ref{subsection-GoLFD}, and present the results in Section \ref{subsection-S-R}.
	\subsection{Generation of Low-Fidelity Data} 
	\label{subsection-GoLFD}
	The advantage of our proposed method lies in its ability to leverage low-fidelity data, e.g., from simulations, to accelerate the optimization process at the high-fidelity level, e.g., using experimental data. To validate this in simulations, fidelity differences are artificially introduced through parameter perturbations of the nominal vehicle model, including $m$ and $l_{\mathrm{f}}$ in \eqref{eq-bicycleModel}, as well as $D_{\mathrm{f,r}}$ in \eqref{eq-mfTire}. Regarding the perturbation of $D_{\mathrm{f,r}}$, $D_{\mathrm{f}}$ and $D_{\mathrm{r}}$ are perturbed simultaneously and will be referred to as $D$ collectively for simplicity. Three simulation studies are conducted, each involving the perturbation of a different parameter. The details of these parameter perturbations are provided in Tab.~\ref{Table-param perturbation}. The perturbations are introduced to influence the closed-loop performance of the control system and do not reflect practical use scenarios of the vehicle. 
	
	To generate the low-fidelity data associated with the perturbed systems, we use BO with a standard (single-fidelity) GP surrogate model for the cost function \eqref{eq-J}. For the generated data set to be sufficiently rich, we employ the integrated posterior variance (IPV) acquisition function in the first 40 runs to globally learn the cost function~\cite{sambu2000} and the EI acquisition function subsequently in the remaining 40 runs to identify potentially optimal regions. This process is conducted once for each perturbed system, yielding three low-fidelity data sets $\mathcal{D}^j_l = (\theta, J_{j}(\theta))$, with $j \in \{1,2,3\}$. The implementation is carried out using BoTorch~\cite{balandat2020botorch}. Eight samples of the cost terms during the generation of the low-fidelity data set are presented in Fig.~\ref{fig-tuning stages}. It can be observed that, as the BO process progresses, the sampled trajectories of the cost terms converge to a similar pattern.
	
	The optimal parameters associated with the smallest costs in the low-fidelity data sets are listed in Tab.~\ref{Table-optima perturbation}. It is evident that for some parameters, the optimal values differ significantly, representing adequate fidelity differences between the nominal and the perturbed systems. This is reasonable because the time-optimal reference trajectory is planned near the friction limit. Hence, perturbations of the model parameters considerably affect the closed-loop performance and introduce inaccuracies in the collected data. 
	\begin{table}[tb]
		\caption{Perturbations of the Nominal Vehicle Model's Parameters.}
		\begin{center}
			\begin{tabular}{ccc}
				\hline
				\textbf{Simulation study} & \textbf{Vehicle parameter} & \textbf{Perturbation} \\ \hline
				\#1 & $m$      & -300kg            \\
				\#2 & $l_{\mathrm{f}}$      & -0.1m            \\
				\#3 & $D$      & $ \times103\%$            \\ \hline
			\end{tabular}
		\end{center}
		\label{Table-param perturbation}
	\end{table}
	\begin{table}[tb]
		\caption{Optimal Controller Parameters of the Nominal and Perturbed Closed-Loop System.}
		\begin{center}
			\begin{tabular}{ccccccc}
				\hline
				& $\bm{\theta_1}$ & $\bm{\theta_2}$ & $\bm{\theta_3}$ & $\bm{\theta_4}$ & $\bm{\theta_5}$ & $\bm{\theta_6}$ \\ \hline
				Nominal  & 0.03      & 0.00       & 0.42       & 0.00       & 0.81       & 0.66       \\
				Sim. \#1 & 0.16      & 0.00       & 1.00       & 0.00       & 0.83       & 0.60       \\
				Sim. \#2 & 0.04      & 0.00       & 0.77       & 0.00       & 0.86       & 0.40       \\
				Sim. \#3 & 0.34      & 0.00       & 0.45       & 0.00       & 1.00       & 0.67       \\ \hline
				\multicolumn{7}{l}{$^{\mathrm{a}}$All parameters are normalized to $[0,1]$.}
			\end{tabular}
		\end{center}
		\label{Table-optima perturbation}
	\end{table}
    
	Given the low-fidelity data, we employ our proposed AR1GP-BO approach as described in Section~\ref{subsection OLCP}. Our proposed approach is further benchmarked against two BO implementations using a NARGP trained on the multi-fidelity data (NARGP-BO) and a single-fidelity GP trained on only the high-fidelity data (SFGP-BO), respectively.
	
	For the implementation, we use the Matérn-5/2 kernel with automatic relevance determination (ARD) to ensure moderate smoothness of the surrogate model~\cite{snoek2012practical}. A Gamma prior distribution $\Gamma(3,6)$ is placed over the kernel lengthscales of the single-fidelity GP as well as the low-fidelity GPs employed in the AR1GP and the NARGP model. All inputs $\theta$ in the data from both fidelity levels are normalized to the range $[0,1]$, and the outputs $J(\theta)$ are standardized to have zero mean and unit variance. 
	The outcomes are averaged over five trials, each of which uses a randomly sampled set of initial parameters. All three methods are implemented using the Python toolkit Emukit~\cite{emukit2023}.
	
\begin{figure*}[bt]
		\centerline{\includegraphics{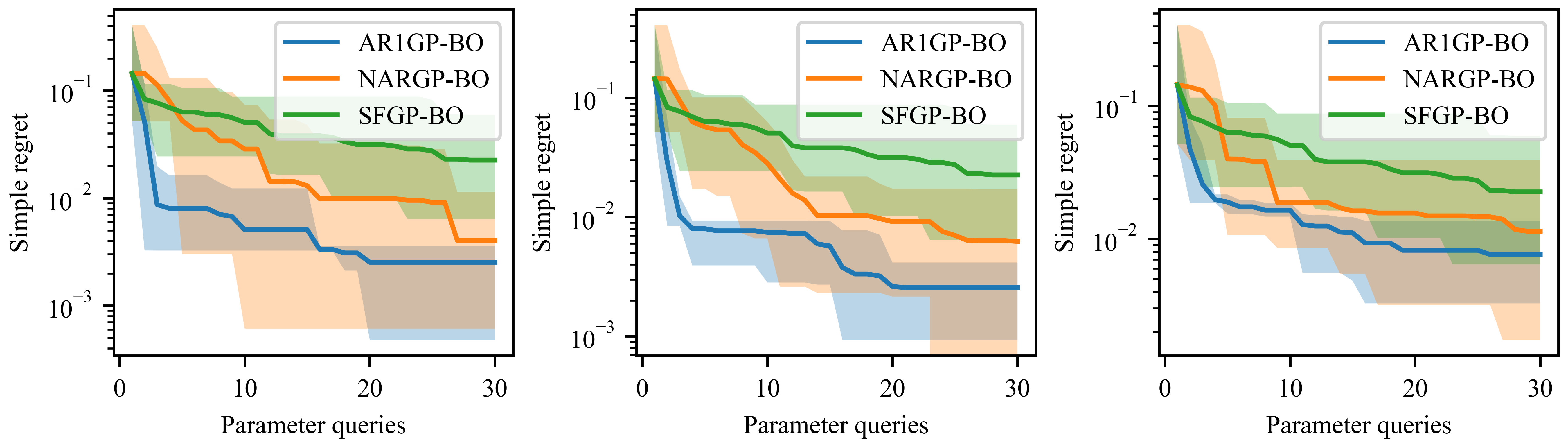}}
    \footnotesize{\qquad\qquad\qquad\qquad\quad(a)~Simulation study \#1. \qquad\qquad\qquad\qquad\qquad(b)~Simulation study \#2. \qquad\qquad\qquad\qquad\qquad(c)~Simulation study \#3.}
		\caption{Comparison of simple regret over the course of the optimization process for our proposed method AR1GP-BO (blue), NARGP-BO (orange), and SFGP-BO (green), averaged over 5 trials each. The shaded areas show the full observed range. Parameter perturbations for generating the low-fidelity data set follow Tab.~\ref{Table-param perturbation}.}
		\label{Fig-benchmark}
	\end{figure*}

	We employ the simple regret value to quantify and compare the performance of the tested methods, which is defined as the difference between the global optimal cost and the best observed cost~\cite{garnett2023bayesian}. For a fair comparison, we present the original values of the regrets rather than the standardized ones. The results are depicted in Fig.~\ref{Fig-benchmark}. Our proposed approach clearly outperforms NARGP-BO and SFGP-BO in terms of convergence speed. Within the first 10 queries, AR1GP-BO converges to a regret level of approximately $10^{-2}$. Besides that, we observe that NARGP-BO outperforms SFGP-BO, which is reasonable since the low-fidelity level adds additional prior information.
	\subsection{Results}
	\label{subsection-S-R} 
	
\begin{figure}[tb]
		\centerline{\includegraphics{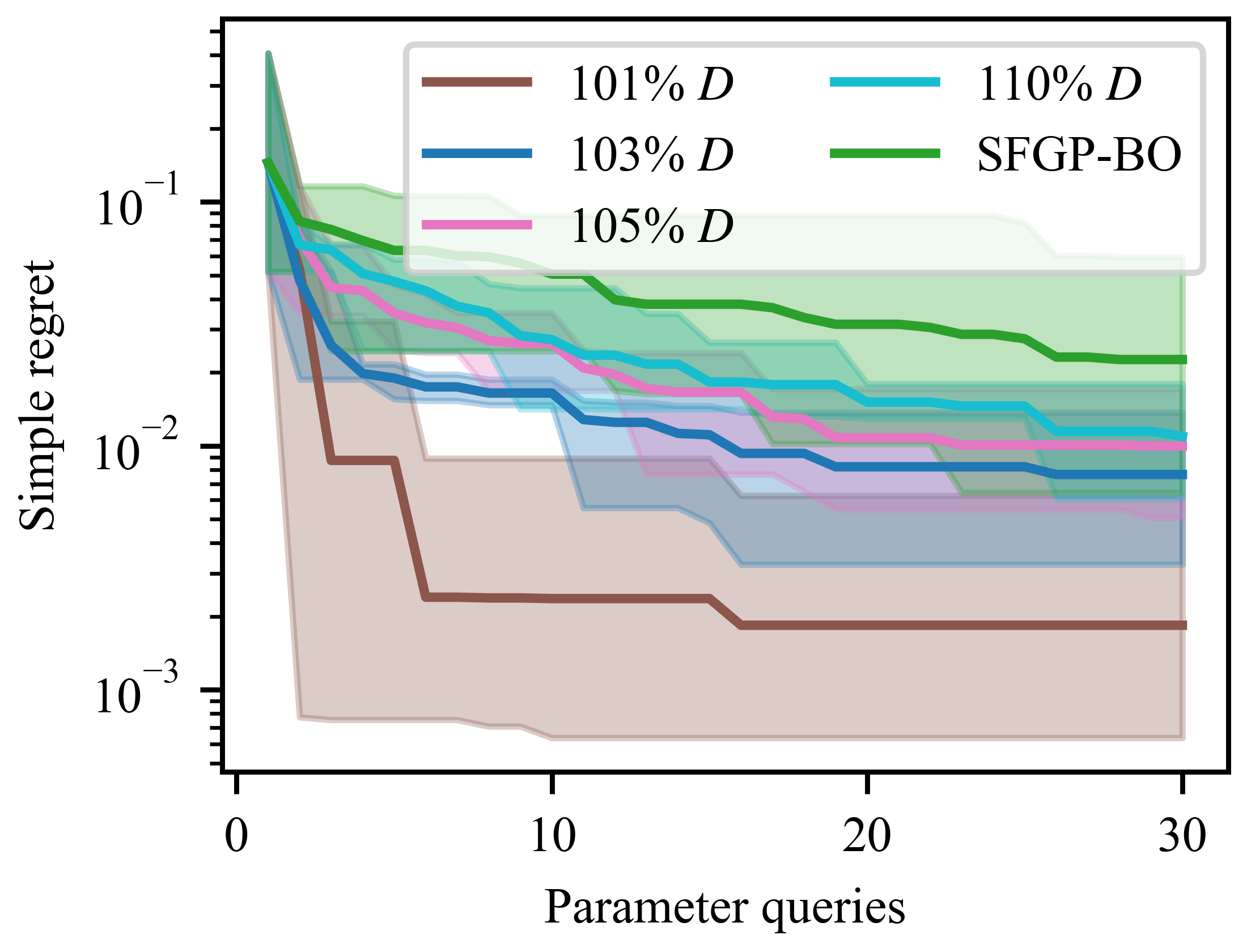}}
		\caption{Influence of the accuracy of the low-fidelity data on the optimization process.}
		\label{Fig-perturb D}
	\end{figure}

	To investigate how the accuracy of the low-fidelity data influences the learning process, we further perturb the parameter $D$ by $1\%$, $5\%$ and $10\%$ of its nominal value. Greater perturbation results in lower accuracy of the low-fidelity data. The results are shown in Fig.~\ref{Fig-perturb D}, indicating that higher accuracy in the low-fidelity data is advantageous for the optimization process. While our proposed approach using the low-fidelity data with $10\%$ perturbation of $D$ still outperforms the SFGP-BO, its convergence rate is noticeably slower compared to cases with less perturbation. 
	
	\section{Real-World Experiments and Validation}\label{Experimental Studies} 
\begin{figure}[bt]
		\centerline{\includegraphics[width=0.55\columnwidth]{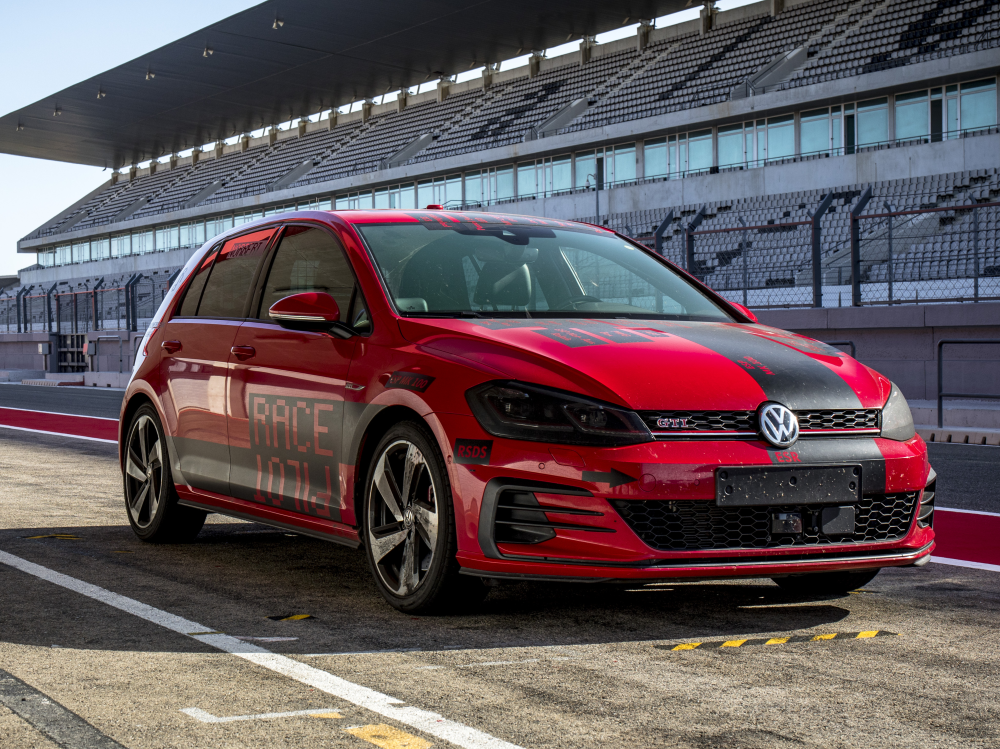}}
		\caption{The test vehicle (Volkswagen Golf VII GTI) with a trajectory tracking control system.}
		\label{Fig-Norbert}
	\end{figure}

	In this section, we present the application of our proposed AR1GP-BO method on a test vehicle (Fig.~\ref{Fig-Norbert}) to learn controller parameters for achieving the optimal closed-loop performance. 
    Low-fidelity data are generated using simulations of the control system in Fig.~\ref{Fig-architecture} and the nominal vehicle model. The sampling strategy described in Section~\ref{subsection-GoLFD} is employed. High-fidelity data are collected through field tests, where the vehicle model depicted in Fig.~\ref{Fig-architecture} is replaced with a test vehicle.
	
    For the configuration of the surrogate model, we adopt the same implementation as described in~\ref{subsection-S-R}. Additionally, a box constraint $[10^{-5},2 \times 10^{-4}]$ is placed over the Gaussian noise variance $\sigma_n^2$. A total of 17 parameter queries are conducted, whereupon the first tested parameter is determined based on only the low-fidelity data set since high-fidelity data are not available in the first run. For controller parameters that do not result in successful completion of the lap, we assign a fixed heuristic cost $J=1$. For comparison, we also evaluated the cost function using the controller parameters manually tuned through field tests and the best parameters in the low-fidelity data set.
	
\begin{figure}[tb]
		\centerline{\includegraphics{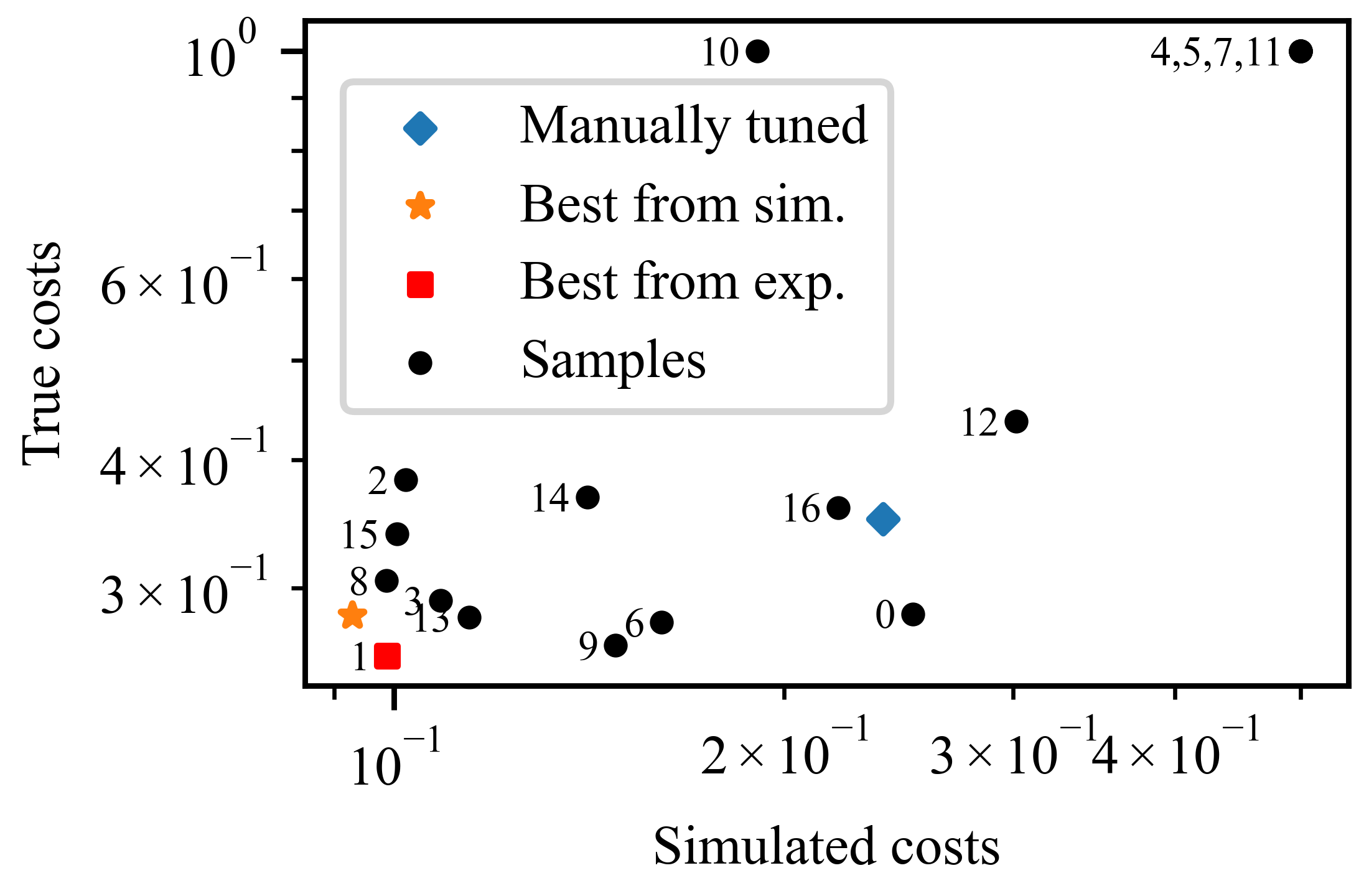}}
		\caption{Simulated and true costs of the sampled controller parameters. The numbers next to the data points indicate the sequence of evaluations.}
		\label{Fig-mismatch}
	\end{figure}

	The true costs of all $17$ controllers sampled during the learning process are depicted in Fig.~\ref{Fig-mismatch} along with their simulated costs. 
    The distribution of the data points clearly indicates the fidelity difference between simulations and field tests. Notably, the best parameter from simulations does not correspond to the lowest true cost. 
    Furthermore, the manually tuned controller parameters are suboptimal in both contexts, which generally suggests that closed-loop performance can be improved through automatic learning of controller parameters with optimization-based approaches.
	
\begin{figure}[bt]
		\centerline{\includegraphics{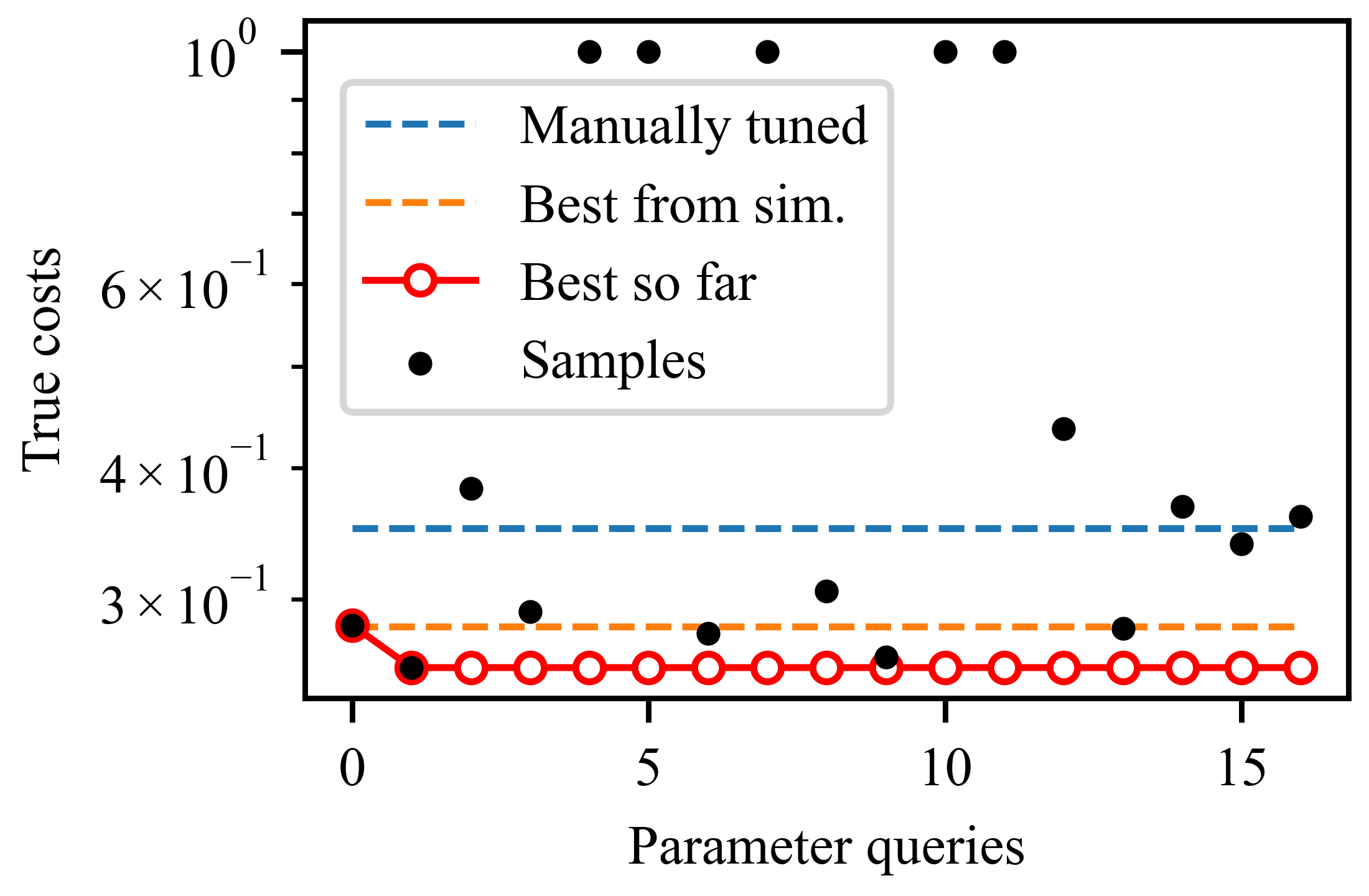}}
		\caption{Cost in each parameter query during the learning process. The blue dashed line represents the true cost of the manually tuned controller parameters, and the orange dashed line indicates the true cost of the optimal parameters acquired from simulations.}
		\label{Fig-experiment}
	\end{figure}

    The evolution of the true cost over the conducted real-world experiments is shown in Fig.~\ref{Fig-experiment}.
    Notably, our proposed AR1GP-BO approach achieves the best observed cost across all tests in the first parameter query. 
    However, given the lack of knowledge about the cost function, it cannot be confirmed whether this cost value represents the global minimum.
    Nevertheless, it already surpasses the cost values obtained with the manually tuned controller parameters and the optimal parameters obtained from simulations.
    The best controller parameter values are provided in Tab.~\ref{Table-optima real}.

    It is worth noting that the controller parameters selected in the $4^\text{th}$, $5^\text{th}$, $7^\text{th}$, $10^\text{th}$, and $11^\text{th}$ run lead to unsuccessful completion of the lap (Fig. \ref{Fig-mismatch} and Fig. \ref{Fig-experiment}).
    While the occurrence of such parameter samples can be explained by the inherent, characteristic trade-off between exploration and exploitation of BO, they are highly detrimental for the learning process. 
    This is because these parameters are far from optimal, and potentially result in unsafe system behavior, and we want the learning algorithm to rather focus on safe regions in the parameter space.
    Given these observations, follow-up works will explore concepts from safe Bayesian optimization~\cite{Berkenkamp2016,Krishnamoorthy2022,Hirt2024a,Hirt2024b,Hirt2025} to avoid unsafe parameter regions.
	
	\section{Conclusions and Outlook}\label{CaO}
    Parameter tuning for vehicle control is a challenging and time-consuming task, as it requires balancing performance, robustness, and adaptability while minimizing the need for costly real-world experiments. In this work, we proposed a multi-fidelity Bayesian optimization approach for learning vehicle controller parameters, effectively reducing the reliance on extensive real-world testing while maintaining high closed-loop performance. By systematically leveraging simulation-based learning and real-world validation, our method improves tuning efficiency and ensures adaptability to varying operating conditions.
    
    We demonstrated the effectiveness of our approach through both simulation studies and real-world experiments on a trajectory tracking controller. The results show that our method outperforms manual tuning and standard Bayesian optimization by efficiently integrating knowledge from multiple fidelity levels, reducing the effort required for real-world testing. Moreover, the approach preserves the conventional two-stage development process, making it practical for industrial applications, where controller tuning is still largely based on manual or heuristic methods.
    
    Beyond the specific case of trajectory tracking, our work highlights the broader potential of Bayesian optimization in vehicle control. The integration of probabilistic models with learning-based optimization opens new possibilities for adaptive, data-efficient, and self-optimizing control strategies. This is particularly relevant for advanced driver-assistance systems and autonomous driving, where traditional parameter tuning methods are impractical or inefficient.
    
    Future research will explore extensions to higher-dimensional parameter spaces, real-time adaptation, and integration with safe learning methods to ensure robust controller tuning under dynamic and uncertain conditions. These findings underscore the broader impact of multi-fidelity learning and Bayesian optimization, paving the way for more automated, efficient, and reliable vehicle control system design.
    \begin{table}[tb]
		\caption{Comparison of Manually Tuned and Learned Controller Parameters.}
		\setlength{\tabcolsep}{4pt} 
		\begin{center}
			\begin{tabular}{cccccccc}
				\hline
				& $\bm{\theta_1}$ & $\bm{\theta_2}$ & $\bm{\theta_3}$ & $\bm{\theta_4}$ & $\bm{\theta_5}$ & $\bm{\theta_6}$ & $\bm{J_{\mathrm{True}}}$ \\ \hline
				Manually tuned  & 0.49      & 0.42       & 0.82       & 0.73       & 0.32       & 0.28      & 0.351 \\
				Best from sim. & 0.03      & 0.00       & 0.42       & 0.00       & 0.81       & 0.66  & 0.282     \\
				Best from exp. & 0.07      & 0.52       & 0.94       & 0.00       & 0.92       & 0.74    & 0.258   \\ \hline
				\multicolumn{8}{l}{$^{\mathrm{a}}$All parameters are normalized to $[0,1]$.} 
			\end{tabular}
		\end{center}
		\label{Table-optima real}
	\end{table}
	\bibliographystyle{IEEEtran}
	\bibliography{IVreference}	
\end{document}